\documentclass[twocolumn,preprintnumbers,amsmath,amssymb,amsfonts,prb,floatfix]{revtex4-1}

\usepackage{graphicx,color}
\usepackage{dcolumn}
\usepackage{bm}

\usepackage{amsmath}

\begin{document}

\title[]{Direct surface charging and alkali-metal doping for tuning the interlayer magnetic order in planar nanostructures}

\author{Tamene R. Dasa
%\footnote{Corresponding author: trdasa@mpi-halle.de} 
and Valeri S. Stepanyuk}

\address{Max-Planck-Institut f\"{u}r Mikrostrukturphysik, Weinberg 2, D-06120 Halle, Germany}

\begin{abstract}

The continuous reduction of magnetic units to ultra small length scales
inspires efforts to look for a suitable means of controlling magnetic states.
In this study we show two surface charge alteration techniques for tuning the interlayer
exchange coupling (IEC) of ferromagnetic layers separated by paramagnetic spacers. Our
$ab-initio$ study reveals that already a modest amount of extra charge can switch the mutual alignment
of the magnetization from anti-ferromagnetic to ferromagnetic or vice verse.
%taking Fe/$\rm Cu(Pt)_N$/Fe trilayer as model system. 
We also propose adsorption of alkali metals as an alternative way of varying the electronic and chemical 
properties of magnetic surfaces. Clear evidence is found that 
the interlayer magnetic order can be reversed by adsorbing alkali metals on the magnetic layer. 
Moreover, alkali metal overlayers strongly enhance the
perpendicular magnetic anisotropy in FePt thin films. These findings
combined with atomistic spin model calculations suggest that electronic or ionic way of surface charging
can have a crucial role for magnetic hardening and spin state control.
\end{abstract}

\maketitle
%\begin{quotation}
%
\section{Introduction}\label{intro}

Efficient ways of manipulating memory units and logic
devices became an emergent motive in the field of spin
electronics \cite{Tsymbal,Spaldin}, in addition to minimizing their sizes. In
the last few decades the reading scheme of the modern
hard disc drive was reduced to the nanoscale level
following the discovery of giant magnetoresistance
effect in metallic multilayers by Grunberg and 
Fert \cite{Binasch1989,Baibich1988}. Current technology uses tunneling 
magnetoresistance for reading the state of magnetic units \cite{Parkin2004,Kryder2009}.
It relies on the relative alignment of the magnetic 
layers interspaced by coupling medium, which results in the
difference of the tunneling resistance. The existence of
such intriguing characteristics is related to the 
delocalized electrons within the coupling medium that are confined.
Apart from the advancement 
in the reading mechanism of the state of the magnetic units, 
the writing head still makes use of magnetic field which has less capability of controlling
magnetic states on short length scales. Alternatively, such magnetic
states can be reversed using the current-induced-spin transfer
torque technique in which the spin-polarized current 
tunnel through the junction and consequently producing
spin-torque \cite{Slonczewski1996,Berger1996}.
It has also been shown that the mutual magnetization in magnetic
tunneling junction can be tuned by applying a voltage
and this effect has similar order of magnitude 
like that of spin transfer torque \cite{Wang2013,Zhu2012,Wu2013}.

Recently a more efficient method has been revealed, on the basis of magnetoelectric coupling,
for tuning the spin alignment in a ferromagnet (FM) 
by both Density Functional theory \cite{Brovko2014,Oda,Fahnel,Ruiz2013,Freeman1} and
experimental studies \cite{Tsymbal,Sahoo2007,Wiesendanger}.
For instance, the anisotropy properties of 2D ferromagnet could be tuned 
either by varying its intrinsic charge carriers \cite{Ruiz2013,Freeman1} or 
the oxidation state by electric means \cite{Bi}. 
On the other, hand external electric field was used in combination with 
multiferroic materials to reverse the relative alignment of ferromagnets
separated by non-magnetic medium (FM/NM/FM) \cite{Pertsev2010,Heron2011,Fechner2012}.
Theoretical studies have shown that the interlayer exchange coupling and
magnetoresistance effect are inter-related \cite{Bruno1992,Zhang1998}.
Nonetheless, direct manipulation of the interlayer exchange coupling 
is one of the feasible and perhaps the least investigated approaches for a
tuning magnetic order. It would even be more compelling to find a possible way
of tuning magnetic coupling or anisotropy in planar nanostructures
with ionic doping. 

Our study verifies a possible route for manipulating
the IEC and the relative alignment of the magnetic
layers interspaced by non-magnetic multilayers, employing
direct surface charging and alkali metal adsorption. The former mechanism relies on 
sequential variation of the intrinsic charge concentration of FM/NM/FM multilayer,
and the results related to this phenomena are presented in Sec.~\ref{surfchg}.
Additionally, in Sec.~\ref{fecu} we propose that the process of 
surface charge alteration can be accomplished by adsorbing
free-electron systems, alkali metals, on surfaces. Indeed, this mechanism alters the
electrochemical features of magnetic surface and leads to reversal of the relative
spin alignment. Apart from tuning the interlayer magnetic order
the magnetic anisotropy of Fe/Pt multilayer can be
significantly enhanced by using alkali metal deposition. To better appreciate such 
increase of the MAE, in Sec.~\ref{fept} the hysteresis loop for FePt multilayer is plotted through the 
atomistic spin model approach. 
The discussion and analysis on these findings is presented in Sec.~\ref{discuse}, followed
by conclusion in Sec.~\ref{conclu}.
%
% -----------------------------------------------Computaional details---------------------------------------------------------------------
%

\section {Computational details}\label{comptech}
 
A thorough $ab-initio$ study was performed within the projector 
augmented wave technique (PAW) \cite{Kresse1999} as implemented in
the VASP code \cite{G. Kresse, Kresse1996}. The local spin density 
approximation (LSDA) is used for the exchange and correlation
interactions \cite{ceperly1980}. Plane wave basis set is used
to describe the Kohn-Sham wave function, and 
in all calculations the plane wave cut off energy of 430~eV was used.
A dense $k$-point mesh of 21x21x1 and smearing width of 2~meV, that are tested for their 
reliability, were employed to obtain a more accurate value of MAE and IEC.
We have used a supercell approach in which
two layers of Fe separated with finite number of Cu spacers
are supported by Pt(001) substrate \cite{ref1}.  
The surface charging effect is introduced by sequentially varying the number of valence
electrons. Additionally, a uniform charge background
is assumed \cite{Makov1995}. Doping of alkali metals is done by capping the top
magnetic layer with two different structures of Na or Li overlayers, namely $p$(1x1) and $c$(2x2).
For the latter configuration a 2x2 supercell is used.
The pairwise exchange coupling or simply the interlayer exchange coupling ($\rm E_{IEC} $)
was calculated as $\rm E_{IEC} = E_{AF} - E_{FM} $,
and for all charge states the spins were aligned along their easy axes.
For 2x2 supercell the interlayer coupling is calculated as,
$\rm E_{IEC} = \frac{1}{4}[E_{AF} - E_{FM} $].
To determine the easy axis of magnetization we have performed a 
fully-relativistic calculation including spin-orbit coupling \cite{Kressesoc}.
The magnetic anisotropy energy (MAE) is evaluated 
by taking the energy difference between two axis of magnetization, 
parallel ([100]) and perpendicular ([001]) to the surface plane.
Similar procedure is followed to calculate the MAE for antiferromagnetic configuration and 
the two spins are oriented anti-parallel to each other for both axis of magnetization.
Here after, we shall denote the charge added to the multilayers (net charge) as $q_{+(-)}$ 
whereas $n$ stands for the charge density of the entire system.

In order to study the magnetization reversal in FePt planar nanostructures,
we use atomistic spin model approach.\cite{vampire}
It is essential enough to investigate this phenomena by mapping the magnetic properties already 
investigated on the basis of electronic structure theory with atomistic spin models. In the latter case the 
energetics of interacting spin with an effective magnetic moment is described by spin Hamiltonian 
which has a form

\begin{equation}
	\begin{split}
	 \label{eq:spin-model}
		%\it{H} & = H_{exc} + H_{MAE} + H_{app} \\
		 \mathbf{H} & = {\sum_{i\neq j}^{}{J_{ij} \mathbf{S}_{i}\cdot \mathbf{S}_{j}}-\it{k}_{u} \sum_{i}^{} (\mathbf{S}_{i}\cdot \mathbf{e})^2} \\
		 & {-\sum_{i}^{} \mu_s\mathbf{S}_{i}\cdot \mathbf{H}_{app} } 
	\end{split}
\end{equation}

where the first, middle and last terms describe the exchange interaction ($H_{exc}$), 
magnetic anisotropy ($H_{MAE}$) and the external magnetic field ($H_{app}$) or Zeeman term respectively. 
$J_{ij}$ is the exchange coupling parameters between atomic sites $i $ and $j $, and the local spin moments are denoted 
with unit vector $\mathbf{S}_{i}$ and $\mathbf{S}_{j}$ obtained from the realistic atomic moments as  $\mathbf{S}_{i} = {m}_s/|{m}_s|$.
The  exchange coupling parameters and the single ion anisotropy, in Eq.~1, are obtained from calculations on the basis
of density functional theory. The magnetization curves are simulated by solving the stochastic Landau-Lifshitz-Gilbert (LLG) equation.
At the atomic level the LLG equation is written as

\begin{equation}
		 \label{eq:LLG}
		 {\frac{\partial \mathbf{S}_{i}}{\partial t}} = -\frac{\gamma}{(1+\lambda^2)} \mathbf{S}_{i}\times
		 [ \mathbf{H}_{i,eff} + \lambda (\mathbf{S}_{i} \times \mathbf{H}_{i,eff})]
\end{equation}

where $\gamma $~=~$ 1.76\times10^{11}$~$\rm T^{-1}s^{-1}$ is the absolute value of the gyromagnetic ratio, $\lambda$ is the damping parameter
and the atomistic effective magnetic field is denoted with $\mathbf{H}_{i,eff}$. The effective magnetic field is derived from the spin Hamiltonian
shown in Eq. 1. \cite{vampire} 
The FePt multilayers are modeled with two Fe layers separated with an effective medium and it extends as 6~nm~$\times$~7~nm within the $xy$ plane.
When simulating the magnetization curve thermal effect and magnetostatic  field are taken into consideration that are 
included in the effective magnetic field \cite{Brown,vampire}. Further discussion on the hysteresis loop of FePt multilayers 
is presented in Sec.~\ref{fept}.
%
% --------------------------------------------------IEC and surface charging-------------------------------------------------------------
%
\section{Switching interlayer exchange coupling with surface charging}\label{surfchg}
First, we present the results concerned with the influence of direct surface charging on the relative
magnetization of two magnetic layers separated with Cu spacer.  
When the charge neutrality of metallic multilayers is perturbed by additional positive or negative charges,
it leads to spatial charge re-distribution mainly towards the surface. Whereas, a reduced variation
of the spatial charge is also observed underneath the surface.
 \begin{figure}[ht]
		\includegraphics{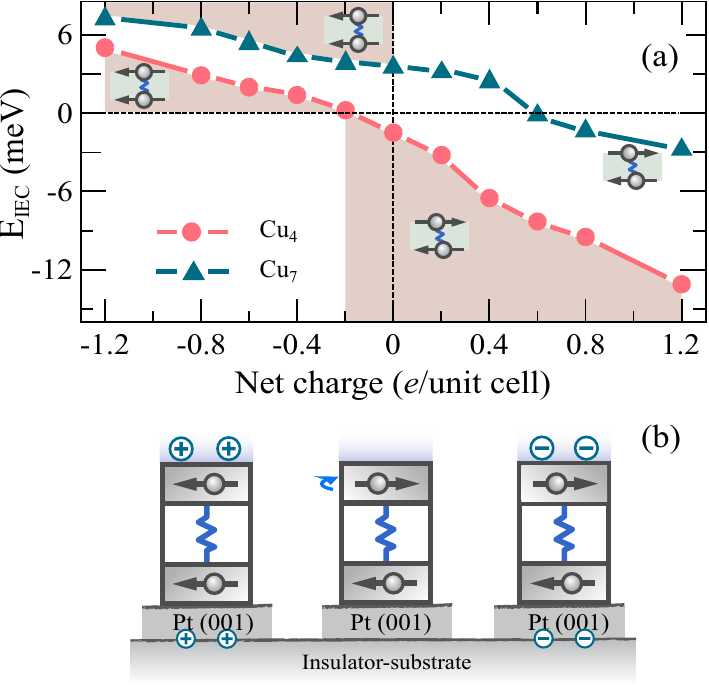}
		\center\caption{(Color online)(a)~The interlayer exchange coupling (IEC) between the magnetic 
		layers in Fe/$\rm Cu_N$/Fe multilayers on Pt(001) as function of the net charge in the system.
		As an illustration such relations are plotted for Cu spacer thickness of N~=~4 (circle) and N~=~7 (triangle). 
		(b)~A schematic description of Fe/$\rm Cu_4$/Fe and the relative 
		alignment of the magnetization as result of charge injection or removal.}
		\label{fig:Fig_1}
\end{figure}
Such charge re-distribution can even be spin-dependent which can alter the magnetic order \cite{Brovko2014, Ruiz2013}.
Meanwhile, the exchange interaction across the coupling medium 
in Fe/$\rm Cu_N$/Fe trilayer, supported on a Pt(001) substrate, determines
the magnetic state to be either ferromagnetic (FM) or
antiferromagnetic (AF). For such multilayers we demonstrate a strong effect
of direct surface charging on the relative magnetic order. As an illustration, 
in Fig.~\ref{fig:Fig_1} the variation of the 
interlayer exchange coupling for Fe/$\rm Cu_N$/Fe multilayers in response to the 
addition or removal of charges is depicted for Cu spacer thickness of N~=~4,~and N~=~7 layers.
The Fe layers separated with four layers of Cu (uncharged)
are coupled antiferomagnetically, and the $\rm E_{IEC}$ is found
to be -2.5 meV. The antiferromagnetic 
coupling can be further enhanced by negative charge doping.
As an example, by injecting negative charge of the order
of  $\Delta $$q$~=~1.2 {\it \={e}} per unit cell~\cite{notechg}  $\rm Fe/Cu_4/Fe$, 
the absolute value of the $\rm E_{IEC}$ is increased to 12 meV.
The most fascinating finding is observed when the electron concentration
is reduced from the neutral $\rm Fe/Cu_4/Fe$, which
results in the switching of the relative magnetic order
from antiferromagnetic coupling to ferromagnetic one.
Explicitly, switching from AF configuration to FM 
takes place when the system is charged with 0.2~$h$ per unit
cell, and further increment of the extra holes to 1.2~$h$
per unit cell leads to $\rm E_{IEC} $ of $\sim5~\rm meV$.

An essential issue now could be the interplay of finite size
effect, $i.e. $ thickness of the Cu spacer, surface charging
and the interlayer exchange coupling. In order to investigate
this phenomena the exchange coupling ($\rm E_{IEC}$) with respect to
charge injection is investigated for thicker Cu layers.
In Fig.~\ref{fig:Fig_1}~(a) we have employed seven layers of
Cu as mediator in which the IEC of the uncharged system is 3.6 ~meV.
Similar to the case of four layers, here space charge variation has significant
effect on the IEC. 
Addition of certain amount of negative charges switches the magnetic coupling to AF.
Interestingly, for $\rm Fe/Cu_7/Fe$ mutilayers a transition 
of the easy axis of magnetization from in-plane to out of plane directions
and the reversal of the relative magnetic order are observed 
at the same charge state, $i.e.$ at 0.6 {\it \={e}}/unit cell.
In another multilayer, $i.e $ $\rm Fe/Cu_6/Fe$ trilayer, we found that 
the magnetic moments are counter-aligned for all charge states. But, the influence
of negative charge injection on the IEC is much stronger than the positive ones.
Precisely, the absolute value of the IEC for the neutral
system ($\sim$6~meV) is reduced by 75~$\%$, as soon as the system is 
injected with 1~{\it \={e}} per unit cell. 
One should note that for the experimental realization of effective surface-interface charging the 
alloy multilayers should be electrically decoupled,
by semiconductor or insulator as shown in Fig.~\ref{fig:Fig_1}~(b), and hence the
net charge is trapped in the upper layers.

%------------------------------------------------------ alkali and magnetic order FeCu -------------------------------------------------

\section{Alkali-metal-induced switching of the interlayer magnetic order} \label{fecu}

\begin{figure}[ht]
		\center\includegraphics{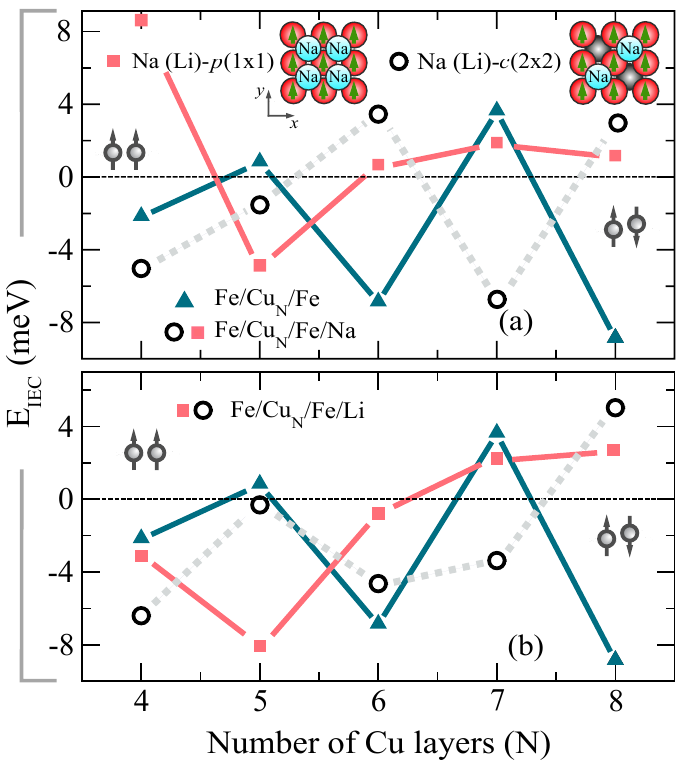}
		\caption{(Color online) (a) Impact of Na adsorption on the interlayer exchange coupling (IEC) 
		in  $\rm Fe/Cu_N/Fe$ (triangles) and $\rm Fe/Cu_N/Fe$/Na- $p$(1x1) (squares) for different
		thickness of Cu spacer. The values of the IEC for $c$(2x2) configuration of Na (top) on the magnetic layer are represented with circles.
		The inset shows the $p$(1x1) and $c$(2x2) structures of Na or Li adsorbates on the magnetic surface.
		(b) Similar plot on the variation of the IEC for $\rm Fe/Cu_N/Fe$ and $\rm Fe/Cu_N/Fe/Li$ (rectangles and circles) multilayers with respect to Cu spacer thickness.  
		The IEC of $\rm Fe/Cu_N/Fe$/Li multilayer with $p$(1x1)[rectangles] and $c$(2x2) [circles] configuration of the Li capping layer is also presented.}
		\label{fig:Fig_2}
	\end{figure}

So far we have shown the impact of charge injection on magnetic order
by controlling the intrinsic charge carriers. Albeit,
controlling the number of charge carriers in a slab could be experimentally challenging task.
Thus, the efforts looking for an alternative method of varying the surface electronic structure are underway. 
Recently it was shown that  incorporating a net charge in nanostructure can be achieved
by electrolyte \cite{Maruyama2009,Weisheit2007,Zhang}, or by adding dopants \cite{Gambardella2013}, 
whereas using scanning tunneling microscope or electron force microscope
tip is also indispensable option \cite{Repp2004}. Doping of magnetic entities with alkali metals or halogens can 
lead to a significant variation of the electrochemical properties \cite{Fratesi2009,Lichtenstein1} as well as their 
spin properties \cite{Cinchetti,Zhao2008}. Moreover, alkali metals are illustrative free-electron systems and have 
relatively small ionization energies. Here we verify that 
alkali-metals deposited on magnetic surface can be employed as a natural way of varying 
the electronic and chemical properties of FM/NM/FM trilayers and the magnetic
properties could be controlled therewith. Particularly,
the interplay between the deposition of alkali metals
and the change in the interlayer magnetic coupling is 
investigated. In Fig.~\ref{fig:Fig_2} (a) the values of the IEC for 
$\rm Fe/Cu_N/Fe$  and Na/$\rm Fe/Cu_N/Fe$ multilayers is presented.
The magnetic coupling of $\rm Fe/Cu_N/Fe$ oscillates between FM and AF order as the 
thickness of the spacer increases, which is related to
the quantum well states (QWS) in the Cu multilayers \cite{Bruno1992,Edwards}. In fact,
the results for six, seven and eight Cu layers 
of the neutral systems are compared with available studies 
based on Korringa-Kohn-Rostoker method \cite{Kowalewski1998}, and the
magnetic order as well as the exchange coupling energies
are in a very good agreement.

We have considered two structures of Na (Li) on the Fe surface, namely 
Na (Li)-$p$(1x1) and Na (Li)-$c$(2x2) structures, in order to have more insight on the influence of
the Na (Li) coverage on the IEC. The former and latter adsorbate structures
covers the magnetic surface fully and partially, respectively. The configuration 
of these two structures on the magnetic surface is shown in Fig.~\ref{fig:Fig_2} (a) (see the inset).
The Na ions are adsorbed at the hollow site which was identified as the most favorable one 
for alkali adsorbates \cite{Fratesi2009,Yamauchi2003}.
Regarding the magnetic coupling, adsorption of Na ions on the magnetic surface brought
an intriguing change to the interlayer exchange coupling.
As shown in Fig.~\ref{fig:Fig_2} (a) (square), when the Fe layer is fully covered with Na [$p$(1x1)]
all the magnetic states of $\rm Fe/Cu_{N}/Fe$ is switched from FM to AF or from AF to FM,
except for N~=~7. Especially a strong change on the IEC, accompanied by magnetic reversal, is observed
when the the Cu spacer thicknesses are four and eight. Reducing the Na layer coverage by half, 
from Na-$p$(1x1) to Na-$c$(2x2), can also lead to significant change of the IEC. 
As shown in Fig.~\ref{fig:Fig_2} (a) (circles), capping the 
$\rm Fe/Cu_{N}/Fe$ multilayer with Na-$c$(2x2) changes the magnetic order from FM
(for the uncapped multilayers) to AF, when the spacer thicknesses are five and seven Cu layers (N~=~5 and N~=~7). Moreover, the 
same configuration of Na varies the interlayer magnetic coupling in $\rm Fe/Cu_{6,8}/Fe$ multilayer from AF to FM one. 
But, covering the top Fe layer of $\rm Fe/Cu_4/Fe$
with sub-monolayer Na-$c$(2x2) does not switch the magnetic order. 
Switching relative magnetic order can be accomplished by increasing the Na coverage, suggesting that
the IEC depends on the percentage coverage of Na overlayer.
Additional manifestation of the effect of Na coverage on the IEC can be seen on
$\rm Fe/Cu_{7}/Fe$ trilayer. In this multilayer when the Na coverage is reduced by 50~$\%$, 
from $p$(1x1) to $c$(2x2), the IEC changes from 2~meV (FM order) 
to -7~meV (AF order), respectively.
These observations lead us to the conclusion that
variations of the IEC induced by Na depends on the configuration
of the adsorbate (percentage coverage) as well as on the spacer thickness.

Similar investigations have been done
by using Li ions as dopant instead of Na [squares in Fig.~\ref{fig:Fig_2}(b)]. 
Covering the $\rm Fe/Cu_N/Fe$ structures with single layer of Li-$p$(1x1) tends 
to reduce the value of IEC for N~=~7 and increases the absolute value of the IEC
for antiferomagnetically coupled $\rm Fe/Cu_4/Fe$.
Moreover, the magnetic order is reversed from FM ($\rm Fe/Cu_5/Fe$) to AF ($\rm Fe/Cu_5/Fe/Li$)
when the spacer thickness is five Cu layers. Conversely,
for $\rm Fe/Cu_{6,8}/Fe$ trilayers Li doping has the tendency of reducing 
the strength of AF coupling, and even the relative magnetic order in  $\rm Fe/Cu_{8}/Fe$
is changed to FM coupling (for $\rm Fe/Cu_{8}/Fe$/Li), as result of Li doping. 
In Fig.~\ref{fig:Fig_2}(b)] we have also presented the IEC for $\rm Fe/Cu_N/Fe/Li$-$c$(2x2)
multilayer, shown in circles. Here reducing the 
percentage coverage of Li ions from $p$(1x1) to $c$(2x2) can only change the 
strength of the exchange coupling. Only for the case of $\rm Fe/Cu_7/Fe$/Li-$p$(1x1) the mutual
magnetic order can be switch to AF coupling by reducing Li coverage [Li-$c$(2x2)] of the magnetic surface.
Interestingly, the change on the IEC of FeCu multilayers that are caused by 
$p$(1x1) configuration of both Na and Li ions are alike. Moreover,
FeCu multilayers that are doped with Na-$c$(2x2) and Li-$c$(2x2) also show
similar exchange coupling in most cases. All in all, except few cases the 
trends in which Li or Na ions change the IEC is analogous to one another, implying that
similar tuning of the relative magnetization in 
FM/NM/FM multilayers can also be accomplished by other alkali metals.
%
%------------------------------------------------------ alkali and magnetic anisotropy of FePt -------------------------------------------------

\section{Magnetic hardening of F\MakeLowercase{e}P\MakeLowercase{t} multilayers with N\MakeLowercase{a} ions}\label{fept}
We have demonstrated a remarkable effect of direct 
surface charging and alkali-metal-doping on the interlayer exchange coupling. 
It is also worthwhile to investigate the impact of Na ions on
the magnetic anisotropy energy (MAE) which determines the stability of magnetic systems.
In this regard materials with FePt composition are known to have high MAE \cite{Oda,Dasa2013,Bruno1}
and are used for the current high density recording media \cite{Klemmer2002}.
Here, we show that MAE of FePt multilayer can be further enhanced by adsorption of Na ions.
In order to verify such way of magnetic hardening effect, we consider 
a model system which consists of two Fe layers separated with Pt multilayers and theses Fe/$\rm Pt_N$/Fe 
stack is supported by Pt(001) substrate. We calculate the MAE as the difference in energy
between two axis of magnetization of the system, $i.e.$ $\rm MAE~=~E_x~-~E_z$.  
Identical calculation is carried out for Fe/$\rm Pt_N$ bilayer on Pt(001), in order to to estimate the 
layer resolved MAE and its relation with the interlayer magnetic coupling \cite{exchangepar}.
	 \begin{figure}[ht]
		\center\includegraphics{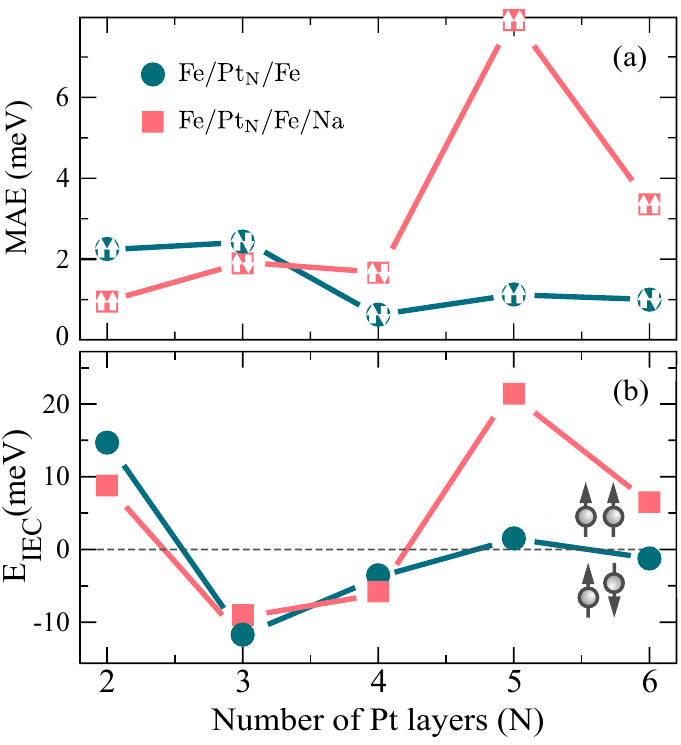}
		\caption{(Color online) (a) The magnetic anisotropy energy (MAE) of the stable magnetic orders in 
		$\rm Fe/Pt_N/Fe$ (circle) and Na/Fe/$\rm Pt_N$/Fe (square) for different
		thickness of Pt spacer is presented. The arrows in the square and circle signs indicate the respective magnetic order
		for which the magnetic anisotropy is evaluated.
		(b) The corresponding plot for the interlayer exchange coupling where positive and negative values of the IEC represent 
		ferromagnetic and antiferromagnetic couplings, respectively.}
		\label{fig:Fig_3}
	\end{figure}

In Fig.~\ref{fig:Fig_3} (a) the MAE of  Fe/$\rm Pt_N$/Fe trilayer is plotted as function of the Pt spacer thickness,
with (squares) and without (circles) the adsorption of Na ions.  
The values of the MAE are presented for the stable magnetic configurations, and in all cases
the easy axis of magnetization is aligned perpendicular to the surface. The FePt multilayers with 
smaller spacer thicknesses (N = 2,3) show high value of MAE relative to the thicker ones.
A decrease in the value
of the MAE of Fe/$\rm Pt_N$/Fe is observed when the Pt spacer thickness increases from N = 3 to N = 4.
Further increment of the spacer thickness brings small change to the magnetic anisotropy and
an average MAE of 1~meV is observed.
As shown in Fig.~\ref{fig:Fig_3} (b) (circles), for Fe/$\rm Pt_{2,5}$/Fe the stable magnetic order is found to be 
ferromagnetic, whereas antiferromagnetic coupling is observed for N~=~3,~4 and 6.
Indeed, the interlayer magnetic order for Fe/$\rm Pt_2$/Fe is in agreement with similar other systems \cite{Wiesendanger2}.
Additionally, experimental and theoretical studies have verified the existence of
such variation of the interlayer exchange coupling between FM and AF coupling in similar systems \cite{Kirschner,Blugel}. 

Basically, Pt is very close to Stoner criterion and can be magnetized when it hybridizes with a ferromagnet.
For example, we found that the nearest Pt layer has a magnetic moment up to 0.35~$\mu_B$, in Fe/$\rm Pt_N$/Fe multilayers.
In such systems the competition of the induced magnetization
from the upper and lower Fe layers determines the magnetic configuration in the neighboring Pt layers,
more importantly for N~$\leq$~3, and probably the magnetic order within the multilayer. As a consequence of such competition the 
Fe/$\rm Pt_{2,3}$/Fe trilayers could be driven away from Stoner instability, leading to antiferomagnetic
order. Certainly, the effect of the spin polarized QWS on the IEC still exits, specially for FePt multilayers with thicker Pt spacer.
Even more such magnetic coupling can be affected by Na capping layer. For instance,
strong change of the IEC is depicted for $\rm Fe/Pt_{5,6}/Fe$, in Fig.~\ref{fig:Fig_3} (b), by Na adsorption. 
Whereas Na capping layer induces moderate change for other FePt multilayers. Additional perspective on the tuning of the IEC in 
relation to the MAE and Na overlayer are pointed out in the latter part of Sec.~\ref{discuse}. 

As already mentioned in the earlier sections one of the main goals of this study
is to point out the impact of alkali metal deposition on the magnetic anisotropy of FePt thin films.
In Fig.~\ref{fig:Fig_3} (a), contrary to the uncapped multilayers, the FePt multilayers
that are capped with Na monolayer show high MAE for thicker spacer thicknesses.
Clearly one can see that Na adsorption strongly enhances
the MAE when the Pt spacer thicknesses of four, five and six atomic layers. 
As an example, deposition of Na ions on Fe/$\rm Pt_6$/Fe trilayers (AF coupling) significantly enhances
the MAE from $\sim$~1~meV to $\sim$~3.4~meV, increasing the energy barrier by more than three fold. 
Exceptionally, for the system whereby the two Fe layers are interspaced with five Pt layer, Na ion 
capping increase the MAE to $\sim$8~meV. Moreover, we have compared the change in the MAE and the IEC and it is clearly seen that
the systems with the strongest IEC corresponds to the highest MAE.
Specifically, we have already shown that Na deposition strongly increases the IEC of Fe/$\rm Pt_5$/Fe trilayers 
from 1.5~meV to 22~meV. The next higher changes in the IEC induced by Na overlayer
are observed when the spacer thicknesses of four and six Pt layers are employed. Likewise, 
the change in the MAE is relatively large. 
These results clearly hint that the enhancement of 
the MAE is almost directly related with the absolute value of the change in the 
interlayer exchange coupling.
Moreover, it verifies that the magnetic order and anisotropy of such complex alloy magnetic systems,
could be altered as result of ionic doping, leading to a change on the magnetic stability
and dynamics \cite{Kimel,Radu}. It is experimentally proven fact that the magnetic order of 
such FM/NM/FM trilayers remains unchanged for thicker ferromagnetic layers \cite{Bloemen,Lang1996},
and hence the interplay between the magnetic order and anisotropy.
	\begin{figure}[ht]
		\center\includegraphics{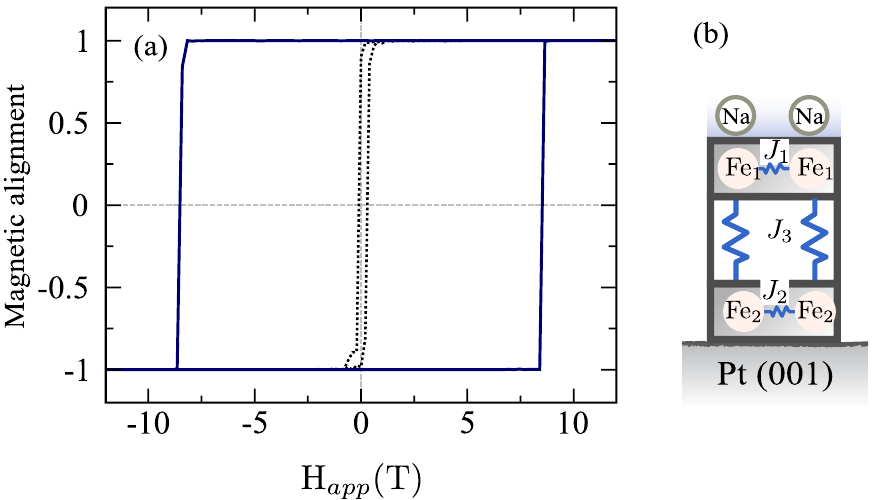}
		\caption{(Color online) (a) The hysteresis loop of Fe/Pt$_5$/Fe trilayers with (full lines) 
		and without (dotted lines) Na capping layers. The simulations are performed for 
		out-of-plane direction which is the easy axis of magnetization. (b) An illustrative
		diagram of the exchange coupling parameters within the multilayer.}
		\label{fig:Fig_4}
	\end{figure}
	
Apart from the magnetic properties discussed above the magnetization reversal process
is studied by simulating the hysteresis loop for FePt multilayers. In order to to do this
we apply an external magnetic field within the range of -12 to 12 T. 
The calculation is performed within the stochastic Landau-Lifshitz-Gilbert (LLG) equation.
The LLG equation (Eq.~2) is integrated by using the Heun integration scheme \cite{Heun}, 
with a time step of $\vartriangle$t = 1 fs. The simulations
temperature is set to be 5~K. Whereas the damping parameters $\lambda$ = 1.0 is used for the 
sake of computational efficiency. 
The values of the exchange coupling parameters, $J_1$, $J_2$ and $J_3$, as well as the
the estimated values of the MAE are presented in TABLE~\ref{tab:parameters} \cite{exchangepar}.
\begin{table}[!ht]
		\caption{The exchange parameters [$J_{1},J_{2}$ and $J_{3} $ (meV/$(\mu_B)^2$)], 
		spin magnetic moments [$m_{\rm_s} $ $\rm Fe_1 (Fe_2) (\mu_B)$] and the estimated MAE (in meV)
		of two Fe layers interspaced with five layers of Pt.}\label{tab:parameters}
	\center\begin{tabular}{c||c|c|c|c|c|c}
     $ $& $J_{1}$ & $J_{2}$ & $J_{3} $ & $\rm MAE_{1}$ & $\rm MAE_{2}$ & $m_{\rm_s} [\rm Fe_1 (Fe_2)]$\\  
     \hline
     \hline
     $\rm Fe_{2}/Pt_{5}/Fe_{1} $     & 5.1 & 4.9 & 0.2 & 0.1 & 1.2 & [2.9 (2.7)]\\
     \hline
     \hline
     $ \rm Fe_{2}/Pt_{5}/Fe_{1}/Na $ & 4.7 & 4.9 & 1.4 & 1.6 & 6.3 & [2.7 (2.9)]\\
      \hline
    \end{tabular}\end{table}
We have created model system of two Fe layers separated with an effective spacer. 
Fig. \ref{fig:Fig_4} illustrates the hysteresis loops for both FePt multilayers with (full lines)
and without (dotted lines) Na deposition. Both multilayers show square like hysteresis loops.
One can see that the Na capping layer
significantly increase the coercive field to H$\rm _a$ = $\sim$8.5~T. Therefore, the atomistic
spin model simulations clearly demonstrate that alkali metal doping of FePt multilayers
favor domain nucleation at elevated temperature by increasing the energy barrier. Furthermore it indicates that 
magnetic hardening effect can be accomplished by Na ion deposition.
%
%------------------------------------------------------final remarks-------------------------------------------------
%
\section{Analysis and discussion of the results}\label{discuse}
\begin{figure}[ht]
		\begin{center}
		\includegraphics[width=8.5cm,scale=6]{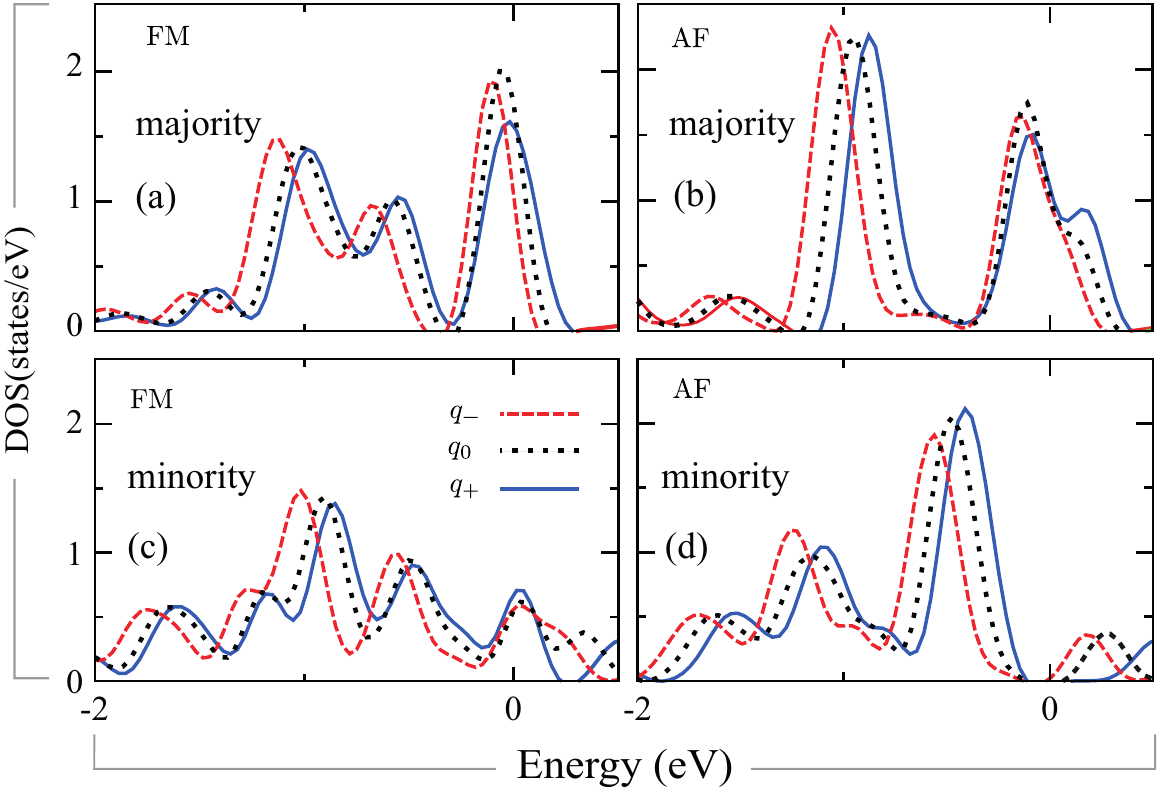}
		\caption{\label{fig:Fig_51} The spin polarized quantum well states of all Cu spacer layers for 
		FM (a and c) and AF (b and d) orders, at different charge states of the 
		supercell are depicted. The majority and minority DOS are presented on the 
		upper (a and b) and lower (c and d) panel, respectively. The density of states are plotted for 
		positive ($q_{+}$=$1~h$, blue full line), negative 
		($q_{-}$=1~{\it \={e}}, red dashed line) and neutral (dotted line)
		$\rm Fe/Cu_4/Fe$ multilayers, close to the center of Brillouin zone.
		The vertical dashed line at zero energy level represents the Fermi energy.}
		\end{center}
	\end{figure}

The findings that have been discussed until now are
supported with analysis on the electronic structure. 
It is known that the interaction between the magnetic layers is mediated by the confined
delocalized electrons in Cu layers \cite{Bruno1992,Brovko2008}, which is a direct 
consequence of the spin-polarized potential at the Fe/Cu interfaces.
Indeed, it is observed that the net charge carriers for negative (positive) charge doping
spills (depletes) mainly at the surface, and minimum variation exists below the surface.
At this stage there are two factors that can impose changes
on the potential of the quantum-well like structure,
where the first one is related to the changes in the 
boundary conditions as a result of the relative alignment of the
two magnetic layers, parallel or anti-parallel. 
Secondly addition or removal of extra charge can alter the Fe/Cu interface
as well as the reflectivity of the confined states at the boundaries \cite{Bruno1992}.
These two modification will affect
the nature of the confinement, being additional 
scattering constraint of the quantum-well states. 
In Fig.~\ref{fig:Fig_51} the effect of these two parameters 
on the spin polarized quantum-well states (QWSs) of $\rm Fe/Cu_4/Fe$ trilayer is
depicted. In these figures the
$s$+$p$ density of states of the coupling medium, for 
all Cu layers, are plotted for FM (a) and AF (b) coupling.
We have considered three charge states, namely positive ($q_{+}$), neutral ($q_{o}$)
and negative ($q_{-}$). An electron or hole doping of the order of
1~{\it \={e}}~($h$) is used for the charged systems. One can infer that 
the shift of the spin polarized QWS in the energy scale depends
on the polarization of the charge doping and consequently
this can alter the nature of the magnetic coupling.
For both magnetic orders the positive (negative) charge doping shifts 
the QWS towards higher (lower) energies. 
on the other hand, the variation of the magnetic order between the magnetic layers can be directly  inferred
from the change in the integrated density of state (DOS) which significantly 
contributes to the total energy \cite{Lang1996,Bruno1992}
, $\rm E = \int^{\varepsilon_F}{{\it n}(\varepsilon)(\varepsilon - \varepsilon_F)d\varepsilon} $, 
where $\rm {\it n}(\varepsilon)$ is the density of states.
Thus, the change in the IEC is explained from the shift in the DOS of the 
magnetic layer or the coupling medium for FM and AF configuration at different charge states.
As shown in Fig.~\ref{fig:Fig_51} the shift
of the QWS induced by positive charging is more enhanced for AF coupling.
The dependence of such band shift on the magnetic order can lead to 
stable ferromagnetic coupling in $\rm Fe/Cu_4/Fe$ for positive charging, 
as it is found in section~\ref{surfchg}. It has to be noted that the QWS are calculated
by sampling the k-points near the center of the Brillouin zone. In order to get additional
information we also analyzed the total density of states.

\begin{figure}[ht]
		\begin{center}
		\includegraphics[width=8.5cm,scale=7]{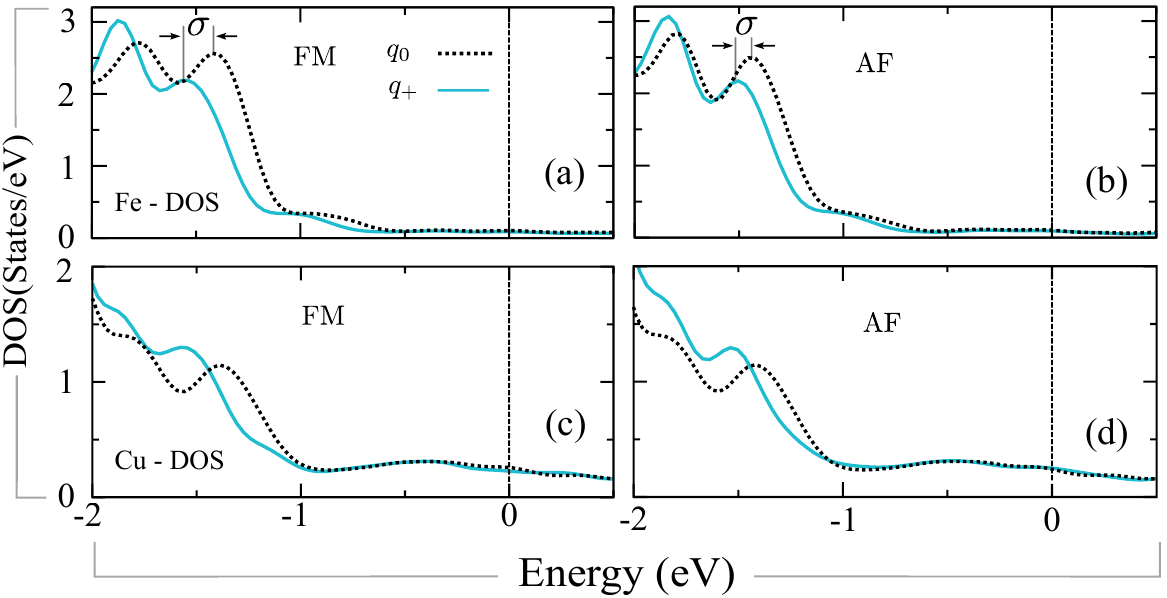}
		\caption{\label{fig:Fig_52} The majority electron density of states (for $d$-orbitals) of the top Fe layer
		$\rm Fe/Cu_4/Fe$, for both FM (a) and AF (b) configurations are shown. The shift in the DOS ($\sigma$) are compared for the neutral 
		system and when it is charged with 1.0~$h$ per unit cell. The majority density of states 
		for the Cu layer interfaced with top Fe layer is shown in the lower panel. The Cu-DOS is plotted
		for the cases where $\rm Fe/Cu_4/Fe$ trilayer is coupled ferromagnetically (c) and antiferomagnetically (b).  
		The density of states are integrated over the whole Brillouin zone.
		The vertical dashed line at zero energy level represents the Fermi energy.}
		\end{center}
	\end{figure}

Studying the variation of the electronic structure of 
the top FeCu interface can also provide more insight in the change of the IEC caused 
by surface charging. 
First, we would like to focus on the electronic structure of the magnetic layer in order to elucidate the 
magnetic reversal scenario once more. 
For both magnetic orders the shift of the minority bands, 
caused by the surface charging process, show minimum difference and are omitted. 
Unlike for the case of Cu-QWS (Fig.~\ref{fig:Fig_51}) the majority and minority bands of the Fe layer
show opposite shift relative to the neutral system. This leads to an increase
or decrease in the exchange splitting as well as the magnetic moment. 
As shown Fig.~\ref{fig:Fig_52}(a) and (b) the majority QWS of Cu spacer
is shifted to higher energy as $\rm Fe/Cu_4/Fe$ trilayer is positively
charged but the majority DOS of Fe are shifted to lower energies.
This is caused by the the net charge accumulated/depleted and the reflectivity of 
the confined electrons within the quantum-well like structure (Fe-Cu-Fe) can be modified
therewith.

In Fig.~\ref{fig:Fig_52} (a) and (b) the majority DOS for $d$-orbitals of the top
Fe layer, for parallel and anti-parallel alignment, at two
different charge states is presented. Similarly, the density of states 
of the Cu atoms that are interfaced with Fe layer is plotted
in Fig.~\ref{fig:Fig_52} (c) and (d). From these density of states one can infer that
there is a hybridization between the $sp$ states of Cu with $d$-orbitals of Fe.
Explicitly, injecting positive charge of 1~$h$ per unit cell induces
a strong shift ($\sigma$) of the majority $d$-bands of Fe towards the
lower energies by $\sigma$~=~140 meV and  $\sigma$~=~65 meV for FM and
AF configurations, respectively. This implies that the
majority bands of Fe, in the case of ferromagnetic coupling,
is more sensitive to positive charge doping which
leads to a lower total energy and stable ferromagnetic
coupling. The reason for such change of the majority
bands is related to the $sp-d$ hybridization of the Cu layer
with top Fe layer. When the system is
either negatively or positively charged the $sp$ states are affected substantially
and it leads to a change in the $sp-d$ hybridization.
Evidently the corresponding shift of the DOS of Cu induced by positive 
charge doping, to the lower energies, has similar tendency like that of the majority bands of 
Fe, see Fig.~\ref{fig:Fig_52} (c) and (d). 

\begin{figure}[ht]
		\center\includegraphics[width=8.5cm,scale=8]{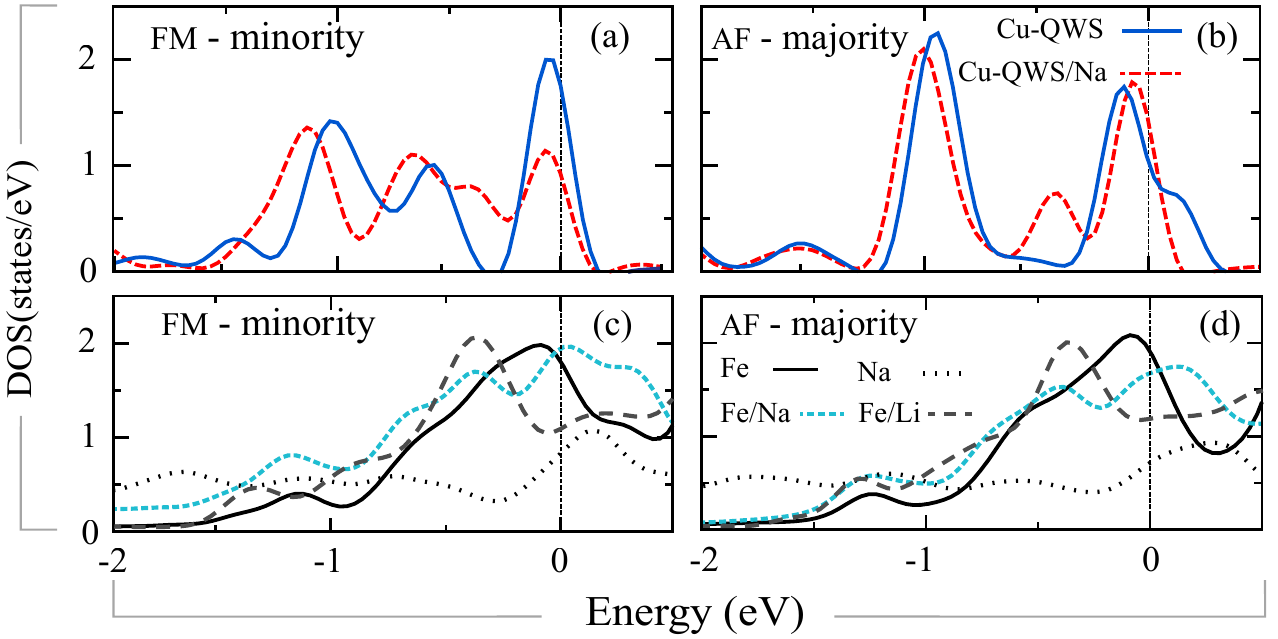}
		\caption{The effect of Na capping layer in the confined states of
		the quantum well like structure, $\rm Fe/Cu_4/Fe$. The DOS
		of the QWS are presented for minority states with [(a), full line]
		and without Na doping [(b), dashed line].
		The minority electronic density of states (for $d$-orbitals) of
		Fe depicted without (full lines) and with (dashed lines) adsorption of Na ions. 
		All the DOSs are plotted for the top Fe layer within the FM (c) and AF (d) configurations of $\rm Fe/Cu_4/Fe$.
		The dashed gray line shows the minority bands of the top Fe in $\rm Fe/Cu_4/Fe$/Li-$p$(1x1) multilayer.
		The DOS for the Na layer is also shown in dotted lines and it is multiplied by factor of ten.
		Since the shift of the majority DOS as result of Na ion doping is relatively similar, 
		for both FM and AF order, and are not included.
		The vertical dashed line at zero energy level represents the Fermi energy.}
		\label{fig:Fig_6}
	\end{figure}

Similar to the case of direct surface charging some insights on the 
variations of the magnetic order (by alkali metal doping) can be inferred
by comparing the QWS and the integrated density of states.
However, the mechanism in which the IEC is changed by 
Na or Li doping involves both chemical and electronic effects, unlike that of direct surface charging. 
In Fig.~\ref{fig:Fig_6} the minority quantum well states within the coupling medium of $\rm Fe/Cu_4/Fe$
are calculated near the center of  Brillouin zone and presented for FM (a) and AF (b) coupling.
In this figure the QWS for both FM and AF orders are displaced to the lower energy as result
of Na doping, which is analogous to the case of negative surface charging. 
Additionally, in Fig.~\ref{fig:Fig_6} the minority DOS of the top Fe layer in $\rm Fe/Cu_4/Fe$ trilayer
is also shown for both FM (c) and AF (d) couplings as well as the DOS of Na capping layers.
In both plots, $i.e.$ upper panel (Cu-QWS) and lower panel (Fe-DOS) the minority bands 
are correspondingly shifted to the lower energy as result of Na capping layer. 
Clearly the effect is more pronounced for the case of FM coupling than the AF ones. 
Following Na adsorption more states are displaced to the lower energies 
in the case of FM coupling than AF one, leading to stable FM order. 
Indeed, similar changes are also observed 
for the bottom magnetic layer which is mediated by quantum well states.
In section~\ref{fecu}, we have seen that either reducing the percentage coverage of Na 
[Na-$c$(2x2)] or Li-$p$(1x1) can reverse the magnetic order of $\rm Fe/Cu_4/Fe$ from AF order to FM order.
To get some insight about these differences we have added the 
minority-DOS plot for top Fe layer [grey dashed line, Fig.~\ref{fig:Fig_6} (c) and (d)] when the
$\rm Fe/Cu_4/Fe$ is doped with Li-$p$(1x1). It seems that Li does not bring 
strong change to the spin polarized electronic structure of the Fe layer which can induce
magnetic reversal. 
According to Fig.~\ref{fig:Fig_6} (c) and (d) the shift of the minority bands 
of Fe to the lower energies by Li-$p$(1x1) overlayer, for FM and AF couplings,
is not as pronounced as in the case of Na-$p$(1x1) doping.
In the contrary, more minority DOS of Fe layers
shifts to lower energy for the case of AF coupling which might only
change the value of the exchange coupling.

Once we have discussed the impact of varying the surface
electronic structure with surface charging or alkali metal doping.
Now, it is important to compare the effects of both surface charging techniques on the IEC.
From Fig.~ \ref{fig:Fig_1} and Fig.~\ref{fig:Fig_2} one can see that
covering the magnetic layers with Na-$c$(1x1) does not necessarily lead to
similar change like that of electron doping (1~{\it \={e}}/unit cell).
As an example, for $\rm Fe/Cu_4/Fe$ trilayer electron doping 
and Na [Na-$c$(1x1)] deposition have opposite effect relative to the neutral system.
However, the impact of the submonolayer Na-$c$(2x2) and the direct surface charging
techniques consistently favors the same magnetic orders for almost all Cu spacer thickness.

Using atomistic spin model simulations, perhaps for the first time we revealed 
magnetic hardening mechanism in FePt multilayers by Na ion doping. 
These variations in the MAE are mainly caused by a combined effect of 
the interlayer exchange coupling and the changes in surface
electronic structure induced by Na. In order to elucidate
the first case we compared the MAE for three different cases, $i.e.$
Fe/$\rm Pt_5$, Fe/$\rm Pt_5$/Fe and Fe/$\rm Pt_5$/Fe/Na on Pt(001). The MAE of 
the first structure is found to be 1.6~meV. When the bilayer Fe/$\rm Pt_5$ is terminated with
Fe layer (Fe/$\rm Pt_5$/Fe), it leads to the onset of the effect of interlayer
exchange coupling on the magnetic anisotropy. The added (top) magnetic layer which couples ferromagnetically
with the bottom Fe layer brings small change to the MAE. However, deposition of
Na ion increases the MAE enormously. Here the strong change of the MAE is the 
ramifications of the IEC and the change in the electronic structure of
the FePt interfaces. This implies that both the magnetic order
and the interface electronic structure are relevant parameters that could possibly
affect the MAE by varying the spin-orbit coupling.
Essentially, the contribution of the spin-orbit coupling to each axis of magnetization, 
in-plane or out-of plane, determines the MAE. For instance, in FePt multilayers the
coupling between occupied and unoccupied states of $d_{\delta}$ and $d_\pi$ orbitals
are identified as the dominant ones among the minority $d-$orbitals \cite{Freeman, Dasa2013},
where $d_{\delta}$~=~$d_{x^2-y^2}$~+~$d_{xy}$ and $d_\pi$~=~$d_{xz(yz)}$~+~$d_{z^2}$.   
For the case of Fe/$\rm Pt_5$/Fe/Na 
the orbitals $d_{\delta}$ ($d_\pi$) are less (more) degenerate near to the Fermi energy 
for the top Fe layer as compared to the lower ones. This tends
to lower the energy when the magnetization of
the multilayer is perpendicular to the surface, leading to a substantial increase in the MAE
for the bottom FePt interfaces. In addition the magnetic moment as well as the induced
magnetic moment for the lower FePt interfaces has increased as result of
Na capping, and conversely reduces for top layers. 

On the other hand some of the FePt multilayers with high MAE
have stable antiferomagnetic coupling,
implying that these FePt thin films could be essential
candidates for AF domains \cite{Nolting}. 
Besides, AF domains have highly minimized dipolar magnetic 
interaction between the neighboring bits,
and could play a crucial role in reducing 
the density of the current magnetic storage
device \cite{Klemmer2002}. Additionally such AF multilayers
can be an integral part of an exchange biased system and 
could have significant impact in the quest 
of ultra-fast magnetic switching \cite{Kimel,Radu}. 
In fact, the switching procedure of the relative magnetization by 
incorporating alkali-metal doping can also be perceived as follows. 
Regardless of the original magnetic order in FM/NM/FM trilayer,
first one has to overcome the energy barrier (MAE) of the top ferromagnet
in order to switch to the opposite direction. The magnetic anisotropy
of each magnetic layer in FePt multilayer is estimated in the earlier section. 
This energy barrier can certainly be altered by introducing alkali metal
in to the system, which varies the interface electronic structure. 
The charge rearrangements are then transformed to magnetic anisotropy through 
spin-orbit coupling which results in changes of the relative orientation 
of the atomic moments. But both surface charging techniques does not always lead to
magnetization reversal. Instead, a favorable conditions will be created for
other means of spin manipulation mechanisms, 
$e.g.$ spin transfer torque \cite{Slonczewski1996,Berger1996} or electrostatic gating \cite{Baure,Chiba,Maruyama2009},
leading to switching mechanism assisted with alkali metal doping.
Therefore, once the energy barrier is completely 
overcome the system probably settles down to the reversed state. 

\section{Conclusion}\label{conclu}

In summary, we have presented a comprehensive study which reveals the effect
of two surface-interface charging techniques on the IEC of
magnetic layers across Cu and Pt spacers as well as their MAE. Explicitly, 
in some cases the mutual
magnetic orders of fully metallic FM/NM/FM-like structure 
can be remarkably tuned between ferromagnetic and
antiferromagnetic states by addition or removal of extra
charge. Moreover, employing alkali metals as dopants
we have shown a natural way of space charge alteration
mechanism which can reverse the relative magnetic orders in 
alloyed multilayers and strongly enhance the MAE of Fe/Pt.
These effect leads us to the conclusion 
that alkali-metal-induced charging magnetic surfaces
can assist the process of magnetization reversal
and also controlling the spin states as well as the spin dynamics.
All in all, our findings suggest that involving alkali metals in field of spintronics 
could have significant impact in order to control and design new functional 
spintronic devices or quantum computing.

\section {Acknowledgment}

T.~R.~Dasa would like to acknowledge Richard F. L. Evans at the University of York for 
the helpful discussion related to the atomistic model approach.
We would also like to thank O.~O.~Brovko and P. Ruiz-D\'{\i}az 
for the fruitful discussion. This work is supported by the
Deutsche Forschungsgemeinschaft (DFG) within the SFB 762.


\begin{thebibliography}{99}
\bibitem{ref1} The layers in the slabs are initially interspaced with a bulk interlayer distance of 1.95~\AA{}, 
 associated with the calculated lattice constant of fcc Pt (3.91~\AA{}). Later on the geometry of the system is 
 relaxed until the force on each atom to the surface is less than 5~meV/\AA{}. The Fe/Cu(Pt)/Fe layers are 
 supported by a substrate which consists of ten Pt layers.
\bibitem{exchangepar} Independent calculation are done in order to find the exchange parameters $J_{1}$ and $J_{2}$ from the supercell
 Pt(001)/Fe/(Na) and Pt(001)/Fe/$\rm Pt_5$ respectively. Based on the number of nearest neighbors ($N_{n}$ ) 
 and the magnetic moments ($m_{i}, m_{j}$) of the interacting spins, the parameters are determined according to $\rm E_{AF} - E_{FM}$ = $2N_{n}J_{ij}m_{i}m_{j}$.
 Regarding the MAE, first we calculate MAE of Pt(001)/Fe/$\rm Pt_5$ and Pt(001)/Fe/$\rm Pt_5$/Fe/Na multilayers, separately.
 Then we took difference in the MAE of these multilayers, the changes in the spin and orbital moments and analysis on the DOS 
 (based on second order perturbation theory\cite{Freeman}) in order to estimate of the MAE of each magnetic layer.
\end{thebibliography}
\end{document}